 \documentclass[final,3p,times,twocolumn,authoryear]{elsarticle}

\usepackage{amsmath}
\usepackage{amssymb}
\usepackage{siunitx}
\usepackage{ulem}

\usepackage{color}
\definecolor{bittersweet}{rgb}{1.0, 0.44, 0.37}
\definecolor{gold}{rgb}{0.83, 0.69, 0.22}
\definecolor{oldmauve}{rgb}{0.4, 0.19, 0.28}
\definecolor{upforestgreen}{rgb}{0.0, 0.27, 0.13}
\definecolor{orange}{rgb}{1.0, 0.5, 0}

\newcommand\refone[1]{#1}
\newcommand\reftwo[1]{#1}
\newcommand\ownupdate[1]{#1}
\newcommand{\vect}[1]{\mathbf{#1}}

\journal{Planetary and Space Science}

\begin{document}

\begin{frontmatter}



\title{}

    \title{Numerical Simulations of Regolith Sampling Processes}
    \author[tue1]{Christoph M.~Sch\"afer}
    \ead{ch.schaefer@uni-tuebingen.de}
    \author[tue1]{Samuel Scherrer}
    \author[airbus]{Robert Buchwald}
    \author[wien]{Thomas I.~Maindl}
    \author[tue2]{Roland Speith}
    \author[tue1]{Wilhelm Kley}

    \address[tue1]{Institut f\"ur Astronomie und Astrophysik, Eberhard-Karls-Universit\"at T\"ubingen, Auf
    der Morgenstelle 10, 72076 T\"ubingen}
    \address[airbus]{Airbus Defence and Space, Airbusallee 1, 28199 Bremen}
    \address[wien]{Department of Astrophysics, University of Vienna, T\"urkenschanzstra{\ss}e 17, 1180
    Vienna}
    \address[tue2]{Physikalisches Institut, Eberhard-Karls-Universit\"at T\"ubingen, Auf der Morgenstelle
    14, 72076 T\"ubingen}

\begin{abstract}
    We present recent improvements in the simulation of regolith sampling processes in microgravity
    using the numerical particle method smooth particle hydrodynamics (SPH). We use an
    elastic-plastic soil constitutive model for large deformation and failure flows for dynamical
    behaviour of regolith.
    In the context of projected small body (asteroid or small moons) sample return missions, we investigate the efficiency and
    feasibility of a particular material sampling method: Brushes sweep material from the asteroid's
    surface into a collecting tray. We analyze the influence of different material parameters of
    regolith such as cohesion and angle of internal friction on the sampling rate. Furthermore, we
    study the sampling process in two environments by varying the surface gravity (Earth's and
    Phobos') and we apply different rotation rates for the brushes.  
    We find good agreement of our sampling simulations on Earth with experiments 
    and provide estimations for the influence of the material properties on the collecting rate.
\end{abstract}


\end{frontmatter}


\section{Introduction}
Regolith is a layer of loose, heterogeneous material. It includes dust, soil, brittle broken rock and
can be found on terrestrial planets, moons and asteroids. 
ESA conducted a feasibility study for a European Phobos sample return mission, which was called Phootprint
\citep{2014cosp...40E1592K, 2014LPICo1795.8030B}, followed by a Phase~A study called Phobos Sample
Return (PhSR), which investigated a joint ESA/Roscosmos mission as well as an ESA standalone scenario. 
Phobos was chosen as a scientifically interesting destination 
due to its unknown formation process and as a unique opportunity to test some of the key components
of a potential follow-up Mars sample return mission. Phobos has a semi-major axis of \SI{9378}{km}
and a sidereal orbital period of \refone{\SI{0.3}{days} in bound rotation}. \refone{Phobos is irregularly
shaped and has dimensions of \num{27} $\times$ \num{22} $\times$ \num{18} \si{\km}}.
Phobos' internal structure is poorly constrained, 
\cite{GRL:GRL26758} provide values for the mean density of about \SI{1876 \pm 20}{\kg \per \metre^3}
and a porosity of 30 \% $\pm$ 5 \%. The surface of Phobos is covered with regolith and the thickness
of the layer varies between \SI{5}{\metre} and \SI{100}{\metre} \citep{Basilevsky201495}.
Various spacecraft missions have investigated the moons of Mars, see the publication by
\cite{Duxbury20149} for a complete review of Russian and American missions. However, no spacecraft
has landed on Phobos yet.

The PhSR study is a technical baseline for a
sample return mission to identify the key technological requirements and address their development
requirements for a sample return mission.
One of the major milestones is the design of a innovative sampling tool installed on the lander
that will allow to collect at least $\SI{100}{g}$ of Phobos' surface material using a novel rotary brush
mechanism. In addition to Phobos' material, a collected sample will presumably
contain pieces of material originated from Mars. 

Computer simulations and computer aided testing become more and more important in the field of
spacecraft design. By the help of modern computers, we can investigate processes under physical
conditions that are not easily available for experiments on Earth, e.g.\ low-gravity fields.
Recently, new models for the dynamic behaviour of regolith have been developed using different
numerical schemes: \cite{Schwartz2014174} use a soft-sphere discrete element method (SSDEM) in the
N-body gravity tree code \texttt{pkdgrav} to investigate low-speed impact simulations into regolith
in support of asteroid sampling mechanisms in the context of the JAXA's Hayabusa and Hayabusa2
missions, and \cite{NAG:NAG688} developed a model for granular flow using SPH in the context of solid
mechanics combined with a Drucker-Prager yield criterion for plastic flow.

The study at hand is performed as part of ESA's PhSR mission and investigates the efficiency of the
rotary brush sampling mechanism by numerical modeling with the latter \ownupdate{SPH} approach. As
part of ESA's Mars Robotic Exploration technology development programme, a prototype of the rotary
brush sampling tool \ownupdate{was also tested under laboratory conditions on Earth and will be tested in
microgravity on a parabolic flight}
\citep{astra2015}.  We \ownupdate{developed} a two-dimensional SPH model for one of the discussed
sampling tool designs and performed several simulations with varying material properties of the
simulated regolith. The main goal was to show the feasibility of the numerical method for the
application of the sampling mechanism and to provide relative values for the sampling rate depending
on the environmental properties. \ownupdate{At first, we will present the physical model for regolith in the
next section, followed by the numerical model and the implementation. In
Section~\ref{section:simulations}, we will show and discuss the simulations of the sampling process.
We will give a conclusion in Section~\ref{section:conclusions}.}

\section{Physical model for regolith\label{sect:physicalmodel}}
The dynamical behaviour of regolith is usually described by a special constitutive model for soil.
 For the simulations carried
out in this study, we applied the elastic-plastic model introduced by \cite{NAG:NAG688}. 
We will provide a short version of their derivation of the model in the following to point out the
differences to our standard model for solid bodies. For the
comprehensive treatise, we refer to their complete analysis in \cite{NAG:NAG688}.
For an
elastic-plastic material, the total strain rate tensor $\dot{\varepsilon}^{\alpha \beta}$, given by
the derivation of the velocity $\vect{v}$ with respect to the coordinate system for small
deformations
\begin{align}
    \dot{\varepsilon}^{\alpha \beta} = \frac{1}{2} \left( \frac{\partial v^\alpha}{\partial x^\beta}
    + \frac{\partial v^\beta}{\partial x^\alpha}\right),
    \label{eq:strainrate}
\end{align}
can be written as the composition of a purely elastic and a totally plastic strain rate tensor
\begin{align}
    \dot{\varepsilon}^{\alpha \beta} =  \dot{\varepsilon}^{\alpha \beta}_\mathrm{e} +
            \dot{\varepsilon}^{\alpha \beta}_\mathrm{p}.
\end{align}
The elastic strain rate tensor $\dot{\varepsilon}^{\alpha \beta}_\mathrm{e}$ is calculated by the
three-dimensional version of Hooke's law
\begin{align}
    \dot{\varepsilon}^{\alpha \beta}_\mathrm{e} = \frac{1}{2\mu} \dot{S}^{\alpha \beta} +
    \frac{1-2\nu}{3E} \dot{\sigma}^{\gamma \gamma} \delta^{\alpha \beta}.
\end{align}
Here, $\dot{S}^{\alpha \beta}$ denotes the deviatoric stress rate tensor, $\dot{\sigma}^{\alpha
\beta}$ is the stress rate tensor, $\mu$, $E$, and $\nu$ are the material dependent parameters shear
modulus, Young's modulus and Poison's ratio, respectively, and the Einstein sum convention is used
($\dot{\sigma}^{\gamma \gamma} = \mathrm{tr}\left({\dot{\sigma}}\right)$).

The plastic strain rate tensor is given by \reftwo{the} following relation of the rate of change of the plastic
multiplier $\lambda$ and the plastic potential function $g$. 
\begin{align}
    \dot{\varepsilon}^{\alpha \beta}_\mathrm{p} = \dot{\lambda} \frac{\partial g}{\partial
    \sigma^{\alpha \beta}}.
\end{align}
If the plastic potential $g$ is equal to
the yield function $f$ of the material, the flow rule is called associated, and otherwise it is
called non-associated. The plastic multiplier $\lambda$ has to satisfy the conditions of the yield
criterion: $\lambda=0$ for $f<0$ or $f=0$ and $\mathrm{d}f <0$, which corresponds to elastic or
plastic unloading and $\lambda>0$ for $f=0$ and $\mathrm{d} f =0$, which corresponds to plastic
loading. 
Following \cite{NAG:NAG688}, the two expressions for the elastic and the plastic strain rate tensors
can be substituted into the equation for the strain rate tensor 
\begin{align}
    \dot{\varepsilon}^{\alpha \beta} = \frac{1}{2\mu} \dot{S}^{\alpha \beta} + \frac{1-2\nu}{3E}
    \dot{\sigma}^{\gamma \gamma} + \dot{\lambda} \frac{\partial g}{\partial \sigma^{\alpha \beta}},
\end{align}
which finally yields an expression for the general stress-strain relationship for an elastic-plastic
material
\begin{align}
    \dot{\sigma}^{\alpha \beta}  = & \  2\mu \left( \dot{\varepsilon}^{\alpha \beta} - \frac{1}{3}
    \dot{\varepsilon}^{\gamma \gamma} \delta^{\alpha \beta} \right)   
     + K \dot{\varepsilon}^{\gamma \gamma} \delta^{\alpha \beta} \nonumber \\ & -
    \lambda \left( \left( K - \frac{2\mu}{3} \right) \frac{\partial g}{\partial \sigma^{\varphi
    \vartheta}} \delta^{\varphi \vartheta} \delta^{\alpha \beta} + 2\mu \frac{\partial g}{\partial
    \sigma^{\alpha \beta}} \right) .
    \label{eq:stressrate}
\end{align}
Here, $K$ denotes the bulk modulus, which is given by the following expression
\begin{align}
    K = \frac{E}{3 - 6\nu}.
\end{align}
In order to calculate the stress in the regolith, one needs an expression for the rate of change of
the plastic multiplier $\dot{\lambda}$. The general formulation for $\dot{\lambda}$ is given by
\begin{align}
    \dot{\lambda} = \frac{ 2 \mu \dot{\varepsilon}^{\alpha \beta} \frac{\partial f}{\partial
    \sigma^{\alpha \beta}} + \left( K - \frac{2\mu}{3} \right) \dot{\varepsilon}^{\gamma \gamma}
    \frac{\partial f}{\partial \sigma^{\alpha \beta}} \delta^{\alpha \beta} }
    {2 \mu \frac{\partial f}{\partial \sigma^{\varphi \vartheta}}\frac{\partial g}{\partial 
    \sigma^{\varphi \vartheta}} + \left( K - \frac{2\mu}{3}\right)\frac{\partial f}{\partial
    \sigma^{\varphi \vartheta}} \delta^{\varphi \vartheta}  \frac{\partial g}{\partial \sigma^{\varphi \vartheta}}
    \delta^{\varphi \vartheta} }. 
    \label{eq:lambdadot}
\end{align}
Hence, as soon as the yield function $f$ and the plastic potential function $g$ are known for a
specific material, the rate of change of the plastic multiplier $\dot{\lambda}$ can be computed, and
consequentially, the stress rate tensor by the use of equation~(\ref{eq:stressrate}). The stress rate
tensor is then integrated and the acceleration due to the stress is determined by the equation for
the conservation of momentum
\begin{align}
    \frac{\mathrm{d} v^\alpha}{\mathrm{d} t} = \frac{1}{\varrho} \frac{\partial \sigma^{\alpha
    \beta}}{\partial x^\beta},
    \label{eq:euler}
\end{align}
where $\varrho$ is the density of the regolith.
The term $k_\mathrm{ext}$ corresponds to a specific volume 
force, e.g., an additional external gravity term.

The evolution of the mass density of the material is governed by the continuity equation
\begin{align}
    \frac{\mathrm{d}\varrho}{\mathrm{d} t} = -\varrho \frac{\partial v^\alpha}{\partial x^\alpha}.
    \label{eq:continuity}
\end{align}
In summary, the set of partial differential
equations~(\ref{eq:stressrate},\ref{eq:continuity},\ref{eq:euler}) is sufficient to describe the
dynamical behaviour of a solid, plastic body. The specific plastic behaviour of the material is given
by the yield function $f$ and the plastic potential function $g$.
The plastic behaviour of granular media can be described by the
model introduced by \cite{DruckerPrager1952}. In the Drucker-Prager model, the yield function is
given by the following relation between the first and second invariants of the stress tensor
\begin{align}
    f(I_1,J_2) = \sqrt{J_2} + \alpha_\phi I_1 - k_c = 0.
\end{align}
The invariants are given by the expressions
\begin{align}
    I_1 = \mathrm{tr}(\sigma) = \sigma^{\gamma \gamma} \qquad \mbox{and} \qquad J_2 = \frac{1}{2} S^{\alpha
    \beta} S_{\alpha \beta}.
\end{align}
The two material constants $\alpha_\phi$ and $k_c$ are called Drucker-Prager's constant and are
related to the Coulomb's material constants cohesion $c$ and angle of internal friction $\phi$.  The
dependence between the four material parameters is different in plane strain and plane stress
conditions. Throughout this study, we focus on plane strain conditions, where the relation is 
given by (see \citealt{NAG:NAG688})
\begin{align}
    \alpha_\phi = \frac{\tan \phi}{\sqrt{9 + 12 \tan^2 \phi}} \qquad \mbox{and}
    \qquad
    k_c = \frac{3c}{\sqrt{9 + 12 \tan^2\phi}}.
\end{align}
In addition to the yield function $f$, the plastic potential function has to be determined to
specify the stress-strain relationship. In the simulations carried out in this project, we use the
non-associated flow rule 
\begin{align}
    g = \sqrt{J_2} + 3 I_1 \sin \psi, \label{eq:plasticflowrule}
\end{align}
where $\psi$ denotes the dilatancy angle. 
We can write the derivative of the plastic potential function with respect to the stress
tensor as
\begin{align}
    \frac{\partial g}{\partial \sigma^{\alpha \beta}} & = \frac{\partial g}{\partial I_1}
    \frac{\partial I_1}{\partial \sigma^{\alpha \beta}} + \frac{\partial g}{\partial \sqrt{J_2}}
    \frac{\partial \sqrt{J_2}}{\partial \sigma^{\alpha \beta}} \nonumber \\
  & = \frac{\partial g}{\partial I_1} \delta^{\alpha \beta} + \frac{1}{2\sqrt{J_2}} \frac{\partial
    g}{\partial \sqrt{J_2}} S^{\alpha \beta}.
\end{align}
In the case of a non-associated plastic flow rule given by equation~(\ref{eq:plasticflowrule}), the
derivative of the plastic potential function reads
\begin{align}
    \frac{\partial g}{\partial \sigma^{\alpha \beta}} = 3 \sin \psi \delta^{\alpha \beta} +
    \frac{1}{2\sqrt{J_2}} S^{\alpha \beta} \label{eq:plasticpotential}
\end{align}
and the derivative of the yield criterion function is given by
\begin{align}
    \frac{\partial f}{\partial \sigma^{\alpha \beta}} = \alpha_\phi \delta^{\alpha \beta} + 
    \frac{1}{2\sqrt{J_2}} S^{\alpha \beta}.
\end{align}
By the use of equations~(\ref{eq:lambdadot}) and (\ref{eq:plasticpotential}), we can finally write the
complete stress-strain relationship as
\begin{align}
    \dot{\sigma}^{\alpha \beta}  = & \  2 \mu \dot{\varepsilon}^{\alpha \beta}
     + \left(K - \frac{2\mu}{3} \right) \mathrm{tr}(\dot{\varepsilon}) \delta^{\alpha \beta} 
     \nonumber \\ &- \dot{\lambda} \left( \left( K - \frac{2\mu}{3} \right) 9 \sin \psi \delta^{\alpha \beta} 
     \right. \nonumber \\ & + 
     2 \mu \left. \left( 3 \sin \psi \delta^{\alpha \beta} + \frac{1}{2\sqrt{J_2}} S^{\alpha \beta} \right)
     \right) \\
      =  & \ \ \  2 \mu \dot{\varepsilon}^{\alpha \beta} + \left(K - \frac{2 \mu}{3} \right)
     \mathrm{tr}(\dot{\varepsilon}) \delta^{\alpha \beta} \nonumber \\ &  - \dot{\lambda} \left( 9K\sin\psi
     \delta^{\alpha \beta} + \frac{\mu}{\sqrt{J_2}} S^{\alpha \beta} \right).
\end{align}
The rate of change of the plastic multiplier is given by
\begin{align}
    \dot{\lambda}
    & = \frac{3K\alpha_\phi \mathrm{tr}(\dot{\varepsilon}) + \frac{\mu}{\sqrt{J_2}} S^{\alpha \beta}
    \dot{\varepsilon}_{\alpha \beta}} { 27 K \alpha_\phi \sin \psi + \mu},
\end{align}
or, if the stress state is not on the yield surface (e.g., $f<0$), $\dot{\lambda} = 0$.
\refone{In the simulations performed throughout of this work, we set the dilatancy angle $\psi$ to zero,
which means there is no volume change of the material due to shearing deformations. For the
non-associated flow rule eq.~(\ref{eq:plasticflowrule}) follows the expression of the potential
function \begin{align} g= \sqrt{J_2}. \end{align}}

In order to have a stress rate that is invariant under rotations of the reference frame, we apply the
Jaumann stress rate (see, e.g.,\ \citealt{gray:2001} and references therein) and write the stress
change with the help of the rotation rate tensor $R$ as 
\begin{align}
    \frac{\mathrm{d}}{\mathrm{d}t} \sigma = \dot{\sigma} - \sigma \cdot R^T - R \cdot \sigma,
\end{align}
where the components of $R$ are calculated according to
\begin{align}
    R^{\alpha \beta} = \frac{1}{2} \left( \frac{\partial v^\alpha}{\partial x^\beta} - \frac{\partial
    v^\beta}{\partial x^\alpha} \right).
\end{align}
Finally, we have the complete set of equations to describe the motion of regolith \refone{for a zero
dilatancy angle}
\begin{align}
    &\frac{\mathrm{d}}{\mathrm{d}t} x^\alpha  = v^\alpha, \\
    & \frac{\mathrm{d}}{\mathrm{d}t} v^\alpha  = \frac{1}{\varrho} \frac{\partial \sigma^{\alpha
    \beta}}{\partial x^\beta} + k_\mathrm{ext}^\alpha,\label{eq:momentum} \\
    & \frac{\mathrm{d}}{\mathrm{d}t} \varrho  = -\varrho \frac{\partial v^\alpha}{\partial x^\alpha}, \\
    &  \frac{\mathrm{d}}{\mathrm{d}t} \sigma^{\alpha \beta} =
    \sigma^{\alpha \gamma} R^{\beta \gamma}  + \sigma^{\gamma \beta} R^{\alpha \gamma} 
    + 2 \mu \dot{\varepsilon}^{\alpha \beta} \nonumber \\ & \qquad \qquad  + \left( K - \frac{2\mu}{3} \right)
    \mathrm{tr}(\dot{\varepsilon} ) \delta^{\alpha \beta} \nonumber  - 
    \refone{\dot{\lambda}  \frac{\mu}{\sqrt{J_2}} S^{\alpha \beta}} ,
    \\
    & \dot{\lambda} = \left\{ \begin{array}{cl} \refone{\frac{3K}{\mu}\alpha_\phi
        \mathrm{tr}(\dot{\varepsilon}) + \frac{1}{\sqrt{J_2}} S^{\alpha \beta}
        \dot{\varepsilon}_{\alpha \beta}}  & \mbox{if } \sqrt{J_2}
        + \alpha_\phi I_1  - k_c < 0, \\
    0 & \mbox{else.} \end{array} \right.
\end{align}
In contrast to the regolith model, we need an additional equation of state (EoS) to model the materials of brushes and
the collecting tray. For these objects, we use the Murnaghan EoS (see, e.g., \citealt{melosh1996impact}), which reads
\begin{align}
    p (\varrho) = \frac{K_0}{n} \left( \frac{\varrho}{\varrho_0} - 1 \right)^n.
\end{align}
Here, $K_0$, $\varrho_0$ and $n$ are material parameters denoting the bulk modulus at zero pressure,
some scaling parameter and the density at zero pressure, respectively. 

\section{Numerical method and implementation}
We have implemented the set of equations from Section~\ref{sect:physicalmodel} in our existing code \texttt{miluphCUDA}
\citep{2016A&A...590A..19S} as a module to simulate regolith within the SPH framework. 

Smooth particle hydrodynamics is a meshless Lagrangian particle method that was first introduced by
\cite{lucy:1977} and \cite{1977MNRAS.181..375G} for the solution of hydrodynamic equations for compressible flows in
astrophysical applications. The SPH method was later extended to model solid bodies.
This work was pioneered by \cite{libersky:1991} with various improvements from the same source later on \citep{randles:1996,
libersky:1997}. The first astrophysical application of SPH with strength of material was by \cite{1994Icar..107...98B}.
Our implementation of the SPH equations in the code mainly follows their work.
For a comprehensive introduction to the basic SPH idea and algorithm, we refer to the excellent publications by
\cite{1990nmns.work..269B} and \cite{1992ARA&A..30..543M}. In this section, we present the regolith
module and SPH equations that we have
implemented in our code for this project. For a complete description of the implementation, we refer
to the code paper \citep{2016A&A...590A..19S}.
The main difference to the usual approach to model solid bodies is the lack of an explicit equation
of state for the regolith material. Instead, the time evolution of the whole stress tensor and not
only the deviatoric stresses is calculated. 

In the following, roman indices $a$, $b$ denote particle indices and the sum in a SPH equation runs over all of the 
interaction partners.

\subsection{SPH equations}
The conservation of mass is given by the continuity equation.
The change of the density of particle $a$ is given by
\begin{equation}
\label{eq:continuity_equation_sph}
\frac{\mathrm{d} \varrho_a}{\mathrm{d} t} =  \varrho_a \sum_b \frac{m_b}{\varrho_b} (v^\alpha_a - v^\alpha_b)
\frac{\partial W_{ab}}{\partial x_\alpha},
\end{equation}
where $W_{ab}$ denotes the kernel function for the interaction of particle $a$ and particle $b$.
Throughout our study, we applied the cubic B-spline kernel introduced by \cite{Monaghan85a}.
The kernel function with the normalisation for two dimensions is written as
\begin{equation}
W(r;h) = \frac{40}{7 \pi h^2} 
\left\{    
\begin{array}{l}  
\left(6(r/h)^3 - 6(r/h)^2 +1 \right) \\ \qquad \qquad \mathrm{for\ } 0 \leq r/h < 1/2 \\ \\ 
2\left(1-r/h\right)^3 \\ \qquad \qquad \mathrm{for\ } 1/2 \leq r/h \leq 1 \\ \\
0   \\ \qquad \qquad   \mathrm{for\ } r/h > 1,
\end{array}
\right. 
\end{equation}
where $r$ is the distance between two interacting particles and $h$ denotes the smoothing length.
The first derivative of the kernel is given by
\begin{equation}
\frac{\partial W(r;h)}{\partial r} = \frac{240}{7\pi h^3} \left\{ 
\begin{array}{l}
3(r/h)^2 - 2(r/h) \\ \qquad \qquad \mathrm{for\ } 0 \leq r/h < 1/2 \\ \\ 
-(1-r/h)^2  \\ \qquad \qquad \mathrm{for\ } 1/2 \leq r/h \leq 1 \\ \\
0   \\ \qquad \qquad   \mathrm{for\ } r/h > 1.
\end{array}
\right.
\end{equation}
In order to calculate the acceleration for particle $a$, we use the 
SPH representation of the equation of motion 
\begin{equation}
\label{eq:accel}
\frac{\mathrm{d} v^\alpha_a}{\mathrm{d}t}  = \sum_b m_b \left[ \frac{ \sigma^{\alpha \beta}_a}{\varrho_a^2}
+ \frac{ \sigma^{\alpha \beta}_b}{\varrho_b^2} \right] \frac{\partial W_{ab}}{\partial x^\beta_b}
    + k_{\mathrm{ext}_a}^\alpha.
\end{equation}
We add some additional artificial viscosity terms to eq.~(\ref{eq:accel}) to prevent particles from unphysical
mutual penetration and to dissipate kinetic energy in the shock front of shock waves.
The artificial viscosity terms were
introduced by \cite{monaghan:1983}. They can be regarded as an additional artificial pressure term in
the equations for the conservation of momentum. The additional pressure is calculated for
each interaction pair $a$ and $b$ as follows: 
\begin{equation} \Pi_{ab} = \frac{-\alpha \overline{c}_{\mathrm{s}_{ab}}  \nu_{ab} + \beta
\nu_{ab}^2}{\overline{\varrho}_{ab}}, \end{equation}
where $\alpha$ and $\beta$ are free parameters that determine the strength of the viscosity, and
$\overline{\varrho}_{ab}$ and $\overline{c}_{\mathrm{s}_{ab}}$ are the averaged quantities for
density and sound speed for the two interacting particles, $\overline{\varrho}_{ab} = (\varrho_a +
\varrho_b) / 2 $ and ${\overline{c}_\mathrm{s}}_{ab} = ({c_\mathrm{s}}_a + {c_\mathrm{s}}_b ) / 2$.
The term $\nu_{ab}$ is an approximation for the divergence and is given by
\begin{equation}
\nu_{ab} = \frac{\overline{h}_{ab} (\vect{v_a} - \vect{v_b} ) \cdot ( \vect{x_a} - \vect{x_b})}{(\vect{x_a} -
\vect{x_b})^2 + \varepsilon_v \overline{h}_{ab}^2}.
\end{equation}
Here, $\overline{h}_{ab}^2$ is the average of the smoothing length of particle $a$ and $b$.
The term $\varepsilon_v \overline{h}_{ab}^2$ in the denominator prevents divergence of $\nu_{ab}$ for particles with
small separations.
This artificial pressure is added to the acceleration of the particle
\begin{equation}
\label{eq:accel_artvisc}
\frac{\mathrm{d} v^\alpha_a}{\mathrm{d}t}  = \sum_b m_b \left[ \frac{ \sigma^{\alpha \beta}_a}{\varrho_a^2}
+ \frac{ \sigma^{\alpha \beta}_b}{\varrho_b^2} - \Pi^\star_{ab} \right ] \frac{\partial W_{ab}}{\partial x^\beta_b}
+ k_\mathrm{ext}^\alpha,
\end{equation}
with $\Pi^\star_{ab} = \Pi_{ab}$ for $(\vect{v_a} - \vect{v_b} ) \cdot ( \vect{x_a} - \vect{x_b}) < 0$ and
$0$ elsewhere. 

Since the model for the regolith does not depend on the internal energy and the Murnaghan EoS is
applied for the solid material, we do not have to solve the energy equation.

In order to determine the evolution of the stress tensor, we calculate the strain rate tensor
according to eq.~(\ref{eq:strainrate}). The SPH representation of the equation reads
\begin{align}
    \dot{\varepsilon}^{\alpha \beta}_a = &
      \frac{1}{2\varrho_a}  
    \sum_b m_b  \left[ \left( v_b^\alpha - v_a^\alpha \right) \frac{\partial W_{ab}}{\partial
    x_a^\beta}\right. 
     + \left. \left( v_b^\beta - v_a^\beta \right) \frac{\partial W_{ab}}{\partial x_a^\alpha} \right].
     \label{eq:strainratesph}
\end{align}
With the use of eq.~(\ref{eq:strainratesph}) we can calculate the rate of change of the plastic
multiplier given by eq.~(\ref{eq:lambdadot}) and finally the evolution of the stress tensor.

We have used \texttt{miluphCUDA}'s default integrator, a Runge-Kutta 2nd order with adaptive
time step, to integrate the evolution of the density, stress tensor, velocities and positions of the
SPH particles. The relative accuracy of the integrator was set to $10^{-6}$.
\subsection{Validation and test run}  In order to validate our implementation of the model by
\cite{NAG:NAG688}, we performed \reftwo{a} comparison simulation of \reftwo{a} two dimensional experiment of
soil collapse. They used small aluminium bars of diameters \SIrange{1}{1.5}{mm}, length \SI{50}{mm} and a density
of $\SI{2650}{\kg \per \m^3}$ to model soil. The bars were initially arranged into a rectangular area
of size $\SI{20}{\cm}\times\SI{10}{\cm}$, surrounded by four flat solid walls on a flat surface.  For
their soil model, they estimated a friction angle of $\SI{19.8}{deg}$, a constant Poisson's ratio of
$\nu=0.3$ and a bulk modulus $K=\SI{700}{kPa}$. The experiment starts by removing one of the four
solid walls, leading to the collapse of the soil.

\begin{figure}
    \begin{center}
        \includegraphics[width=0.5\textwidth]{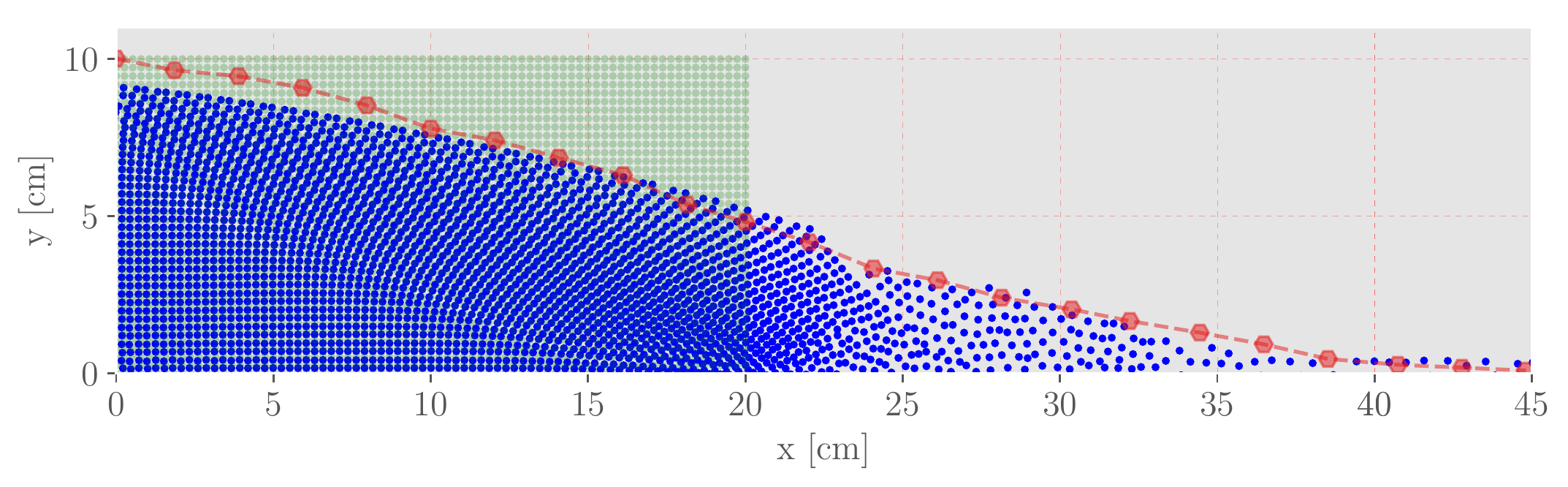}
    \end{center}
    \caption{Result of the validation comparison run, a soil collapse simulation. The blue points
    represent the locations of the SPH particles at the end of the simulation after $\SI{1}{\s}$ has passed, the green points
    indicate the initial positions of the soil ($\SI{20}{\cm} \times \SI{10}{\cm}$). \refone{The red
    connected dots are data points from the cliff collapse experiment by \cite{NAG:NAG688}.} \label{fig:testrun}}
\end{figure}

The parameters for the artificial viscosity were $\alpha=0.3$, $\beta=0.7$, a fixed sound speed of
$c_s=\SI{600}{\metre \per \s}$ and no-slip boundary conditions were applied. The initial number of
particles was $2926$ ($77 \times 38$), the initial particle separation was
\ownupdate{$\Delta=\SI{2.582e-03}{m}$} and the smoothing length was set to $2.5\times \Delta$.  The initial
and final locations of the SPH particles are shown in Figure~\ref{fig:testrun}.  \refone{The plot
additionally shows the outcome from the cliff collapse experiment by \cite{NAG:NAG688} as red
connected data points. The simulation successfully reproduces the shape and length of the collapsed
cliff.}

\section{Numerical simulations of the sampling process\label{section:simulations}}
\ownupdate{In this section, we present and discuss the simulations regarding the sampling process
with the device prototype.}
\subsection{Investigated parameter space}
\begin{table*}
    \begin{center}
    \begin{tabular}{lcccc}
        Gravity $\mathrm{g}$ $\left[\si{m \per s^2}\right]$& Motor [RPM] & Angle of friction [deg]& Cohesion [kPa]
        & Run Identifier \\
        \hline \hline
        9.81 (Earth) & 100 & 42 & 0 & E0 \\ 
        9.81 & 100 & 42 & 0.5 & E1 \\
        9.81 & 300 & 42 & 0  & E2 \\
        9.81 & 500 & 42 & 0 & E3  \\
        \hline
        \num{5.7e-3} (Phobos) & 100 & 42 &  0 & P0 \\
        \num{5.7e-3} & 100 & 42 &  0.5 & P1  \\
        \num{5.7e-3} & 100 & 20 &  1 & P2 \\
        \num{5.7e-3} & 100 & 42 &  1 & P3 \\
        \num{5.7e-3} & 100 & 60 &  1 & P4 \\
        \num{5.7e-3} & 500 & 42 &  0 & P5 \\
        \num{5.7e-3} & 500 & 42 &  1 & P6 \\
        \num{5.7e-3} & -100 & 42 &  0 & P7 \\
        \num{5.7e-3} & -100 & 42 &  0.1 & P8 
    \end{tabular}
    \end{center}
\caption{Material and setup parameters for the sampling simulations. The rotation speed of the motor
    is positive in the collecting phase and negative for the clearing phase by definition.
    \label{table:parameterspace}}
\end{table*}

We have chosen the material properties for the simulations to fit into the range given by
\cite{astra2015} for the regolith simulant that they used for the experiments.
The material parameters for regolith that remain unchanged for all simulations are bulk modulus $\SI{3.33}{MPa}$, shear modulus
$\SI{370}{kPa}$, and bulk density $\SI{1.5}{g \per cm^3}$. We have varied the angle of internal
friction and cohesion. An overview of all simulations performed for this study is provided in Table~\ref{table:parameterspace}. 

\subsection{Initial setup, model for the brushes and the collecting tray}
\begin{figure}
    \begin{center}
        \includegraphics[width=0.4\textwidth]{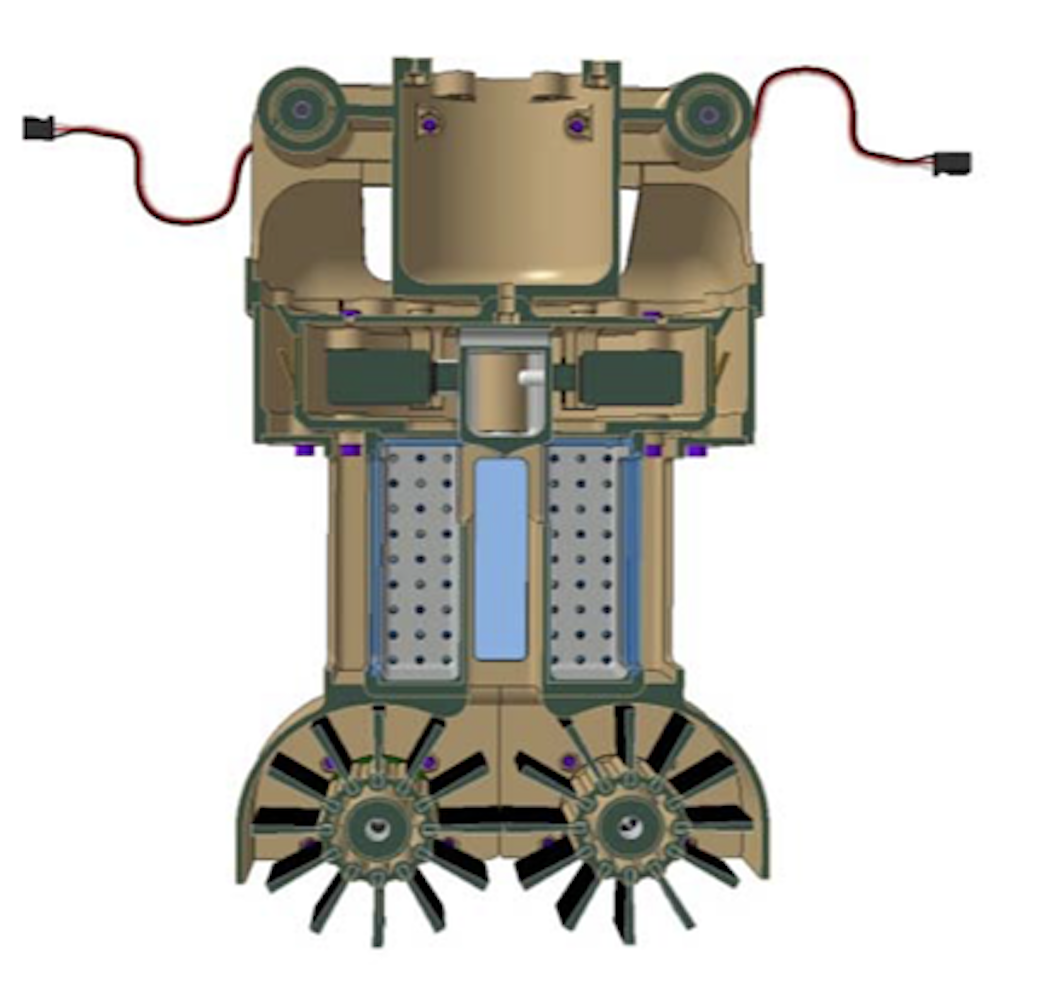}
    \end{center}
\caption{CAD Model of the prototype sampling tool, which acted as our model for the initial SPH
setup.
\label{fig:cad_model}}
\end{figure}

In order to raise the spatial resolution, we have developed a two dimensional model of the existing
CAD data of the sampling tool (see Figure~\ref{fig:cad_model}).

\begin{figure}
    \begin{center}
        \includegraphics[width=0.4\textwidth]{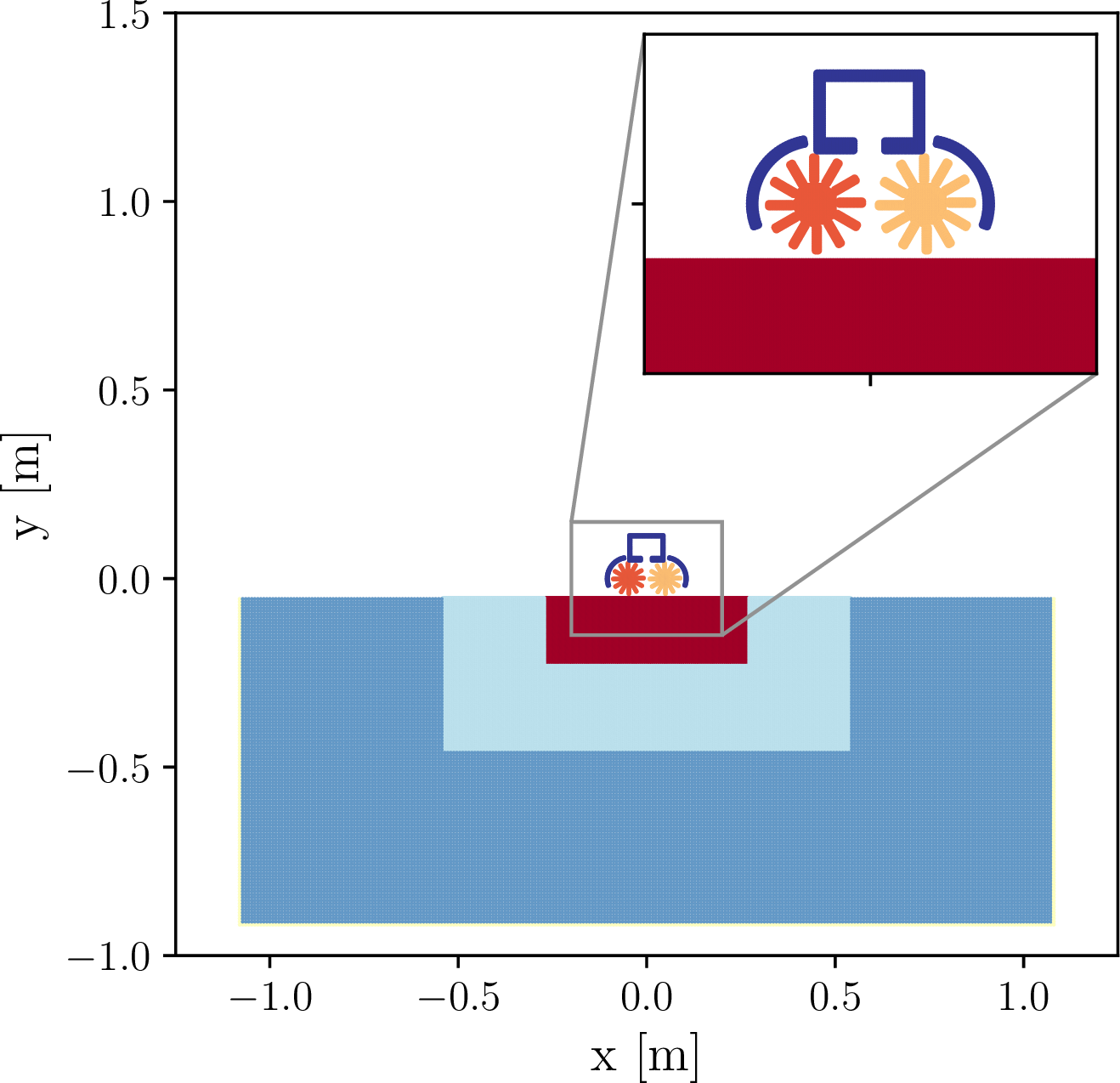}
    \end{center}
    \caption{The initial setup of the sampling simulations. Regolith is colored red, light and dark
    blue, the two brushes
    are orange and yellow, respectively, and the collecting tray and the shielding plates are dark blue.
    Light and dark blue regolith regions are needed to damp sound waves from the inner region of
    interest. In these areas the particle separation is larger and the particles are more massive.
    The outermost layer of boundary particles is fixed. The damping regions are chosen to be large
    enough to suppress interfering with reflections at the outer boundary layer.  \label{fig:setup}}
\end{figure}

The initial setup and geometry of the sampling tool are shown in Figure~\ref{fig:setup}. The
collecting tray is located above two brushes. The brushes ought to shovel regolith from the surface
into the tray. The collecting tray has a height of \SI{63}{mm} and a width of \SI{82}{mm}. On the
bottom of the tray is a opening with a diameter of \SI{26}{mm}. 
The final sampling tool CAD design includes a closure for the opening at the bottom of the collecting
tray.

The two brushes are modelled with 12
individual bristle, separated by \SI{30}{deg}. The inner part of a brush is a cylinder with a
radius of \SI{15}{mm}, the bristles sit on top of this inner part and have a length of \SI{20}{mm}
each. On the outer part of each brush is a circular shielding, covering an opening angle of
\SI{80}{deg}. The purpose of the shielding is to lower the contamination of parts of the landing
device during the sampling process and reduce staining of cameras and sensor devices. The distance between the
pivot point of the brushes is \SI{95}{mm}. 

In the collecting phase, the orange brush rotates counterclockwise and the yellow one
clockwise. During the clearing phase, both brushes rotate in the opposite direction. 
We have applied the Murnaghan EoS for the brushes and the collecting tray including the shielding
with the following material parameters: bulk modulus $\SI{24.2}{GPa}$, bulk density at zero pressure
$\SI{2.7}{g \per cm^3}$, and Murnaghan parameter $n=6$.

Initially, the lowest part of the brush is \ownupdate{\SI{8.4}{\mm}} above the surface. Due to the spatial
resolution given by the smoothing length, the brush touches the surface at a distance of
\ownupdate{\SI{3}{\mm}}.
The brushes and the collecting tray move at constant speed of $v_y= \SI{-5}{\cm \per \s}$ during the
first $\SI{0.2}{\s}$ after simulation start and remain at their final location after this time until
the end of the simulation. Additionally, the brushes start rotating from the beginning and their
rotation speed is fixed. The total simulation time is \SIrange{1}{2}{\s}. The number of particles is
approximately $\num{280000}$ and the smoothing length is set to \SI{3}{mm} (for the red region of
the regolith) in each run.

In order to analyze the efficiency of the collecting process and the dependency on the material
properties of regolith, we monitor the number of particles that get captured in the collecting tray
during the collection phase. This study does not aim to provide absolute values for collecting rates,
we rather investigate the relative difference and the influence of material properties on the
collection process. Hence, the number of collected particles in the tray is a sufficient indicator
for the quality of the sampling rate for the purpose of this study.

%

\ownupdate{In the remaining parts of this section, we present the results of 13 selected simulations of the
sampling process on Earth and Phobos. We will discuss the influence of particular material parameters
such as cohesion and angle of internal friction (see Table~\ref{table:parameterspace} for the
parameters in each run) on the collection rate.}

\ownupdate{
\subsection{Sampling on Earth}
}

\begin{figure*}
    \begin{center}
        \includegraphics[width=0.8\textwidth]{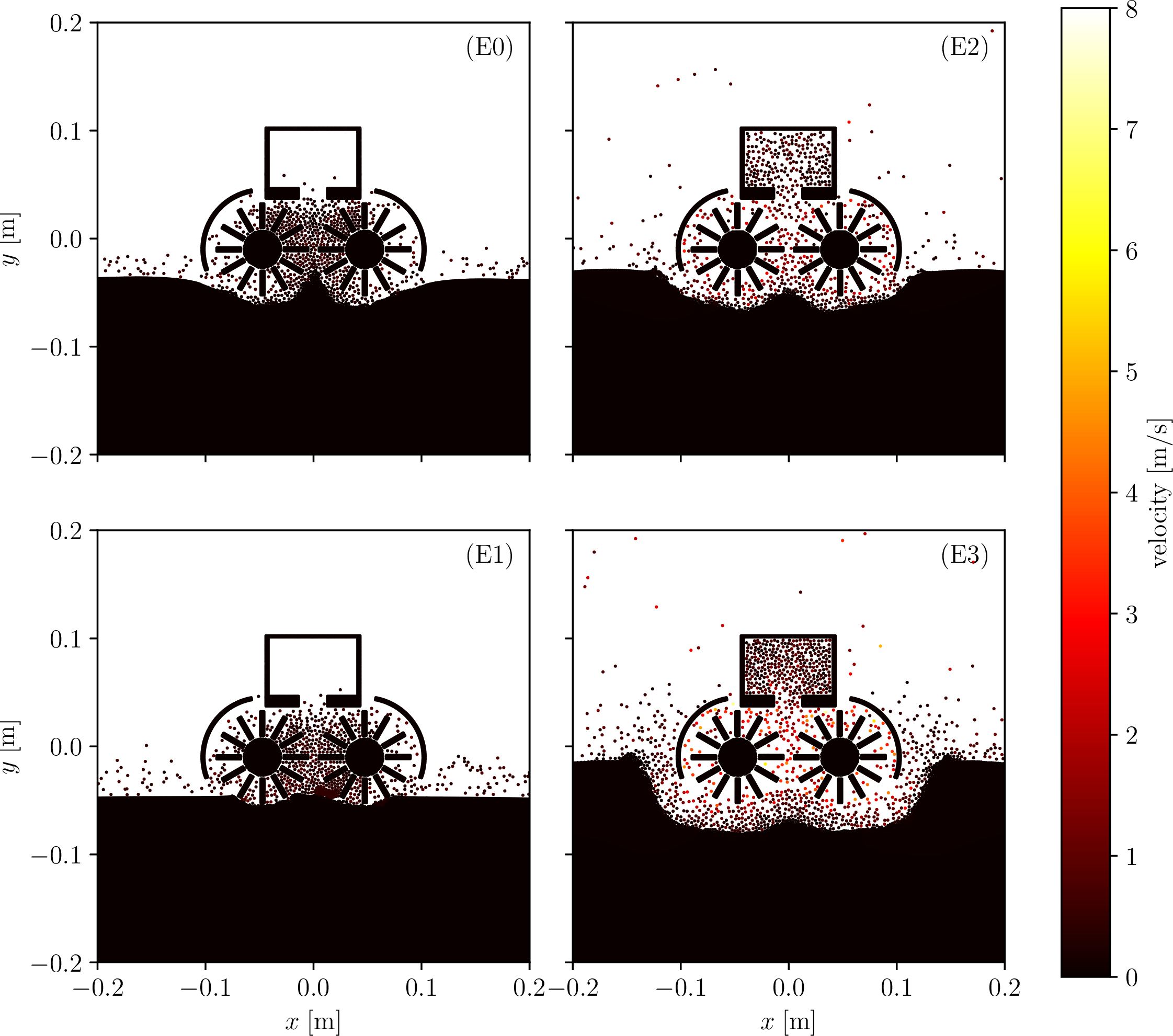}
    \end{center}
    \caption{\label{fig:earth}Particle distribution after \SI{1}{\s} simulation time for four runs with Earth's
    gravity. The velocity of the particles is color-coded. \reftwo{The angle of internal friction is
    \SI{42}{deg} for all runs, the rotation speed of the brushes is \SI{100}{rpm} for runs E0 and
    E1, \SI{300}{rpm} for run E2 and \SI{500}{rpm} for run E3. All runs except E1 with a cohesion of
    \SI{500}{\Pa} are cohesionless. }} 
\end{figure*}
\begin{figure}
    \begin{center}
        \includegraphics[width=0.5\textwidth]{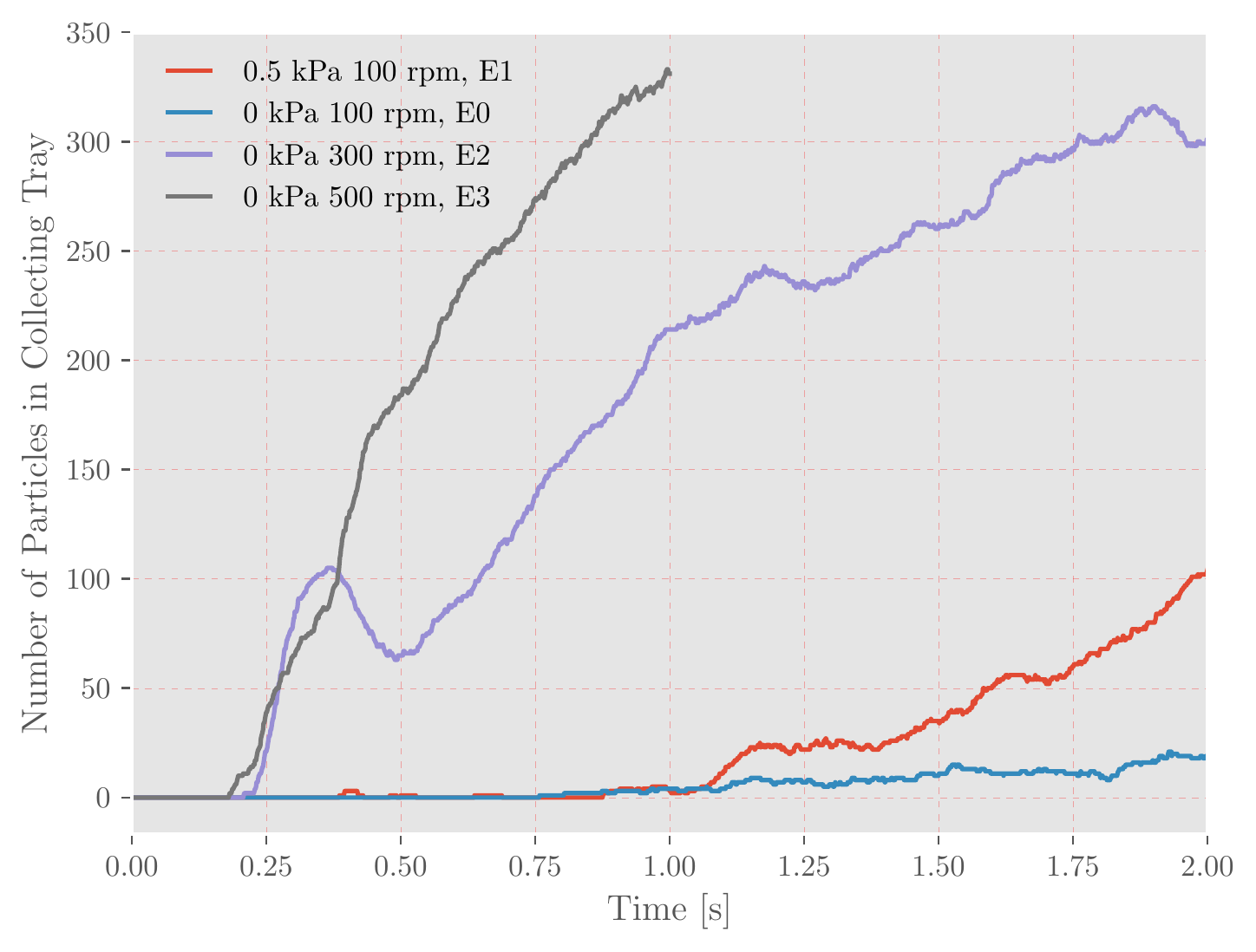}
    \end{center}
    \caption{Sampling process on Earth's surface for regolith with an angle of internal friction of
    \SI{42}{deg}. On Earth, a rotation speed of at least \SI{300}{rpm} is required to transfer regolith
    from the surface into the collecting tray if the brushes run for a shorter time than $\SI{1}{\s}$. 
\label{fig:dust_analysis_earth42}}
\end{figure}

\ownupdate{ We have started our investigations with the sampling process under Earth's gravity
conditions to compare with the experiments performed by \cite{astra2015}.
They conducted a test campaign with an early breadboard design to narrow down bristle material,
configuration, motor torque and rotation speed. In total, more than 300 collecting sequences were
performed. }

\ownupdate{The minimum required rotation speed to collect material in Earth's gravity field is
investigated by performing four different simulations with three different rotation speeds.
Figure~\ref{fig:earth} shows the particle distribution after \SI{1}{\s} simulation time for these
runs. The colour code indicates the speed of the particles. The runs E0, E2 and E3 use cohesionless
regolith with the same angle of internal friction of \SI{42}{deg} for three different rotation speeds
of the brushes, \SIlist{100;300;500}{rpm}. Run E1 has the same parameters as E0 but a cohesion of
\SI{500}{\Pa}.  The particle distribution in between the bristles for runs E0 and E1 and the
corresponding speed of the particles are alike. In the cohesionless case, the brushes have already
left deeper traces on the surface. In both runs, the collecting tray remains empty, only single
particles stray in. Runs E2 and E3 with the higher rotation speeds of \SI{300}{rpm} and \SI{500}{rpm}
display higher velocities of the particles. Additionally, the impact on the surface morphology raises
with the rotation speed: E3 with the highest rotation speed reveals a cavity almost \SI{20}{\cm} in
diameter. While some parts of the material from the cavity were transported to the collecting tray,
some parts were also pushed sidewards. In both runs with the higher rotation speed, the sampling
process works effectively and material from the surface gets lifted into the tray. }

\ownupdate{To analyze the collecting process quantitavely, the number of particles in the
collecting tray during the simulation time is tracked and plotted in Figure~\ref{fig:dust_analysis_earth42}.  A
rotation speed of the brushes of \SI{100}{rpm} is not enough to transfer material into the collecting
tray during the first second (run E0). Adding a little bit of cohesion to the regolith enhances the
sticking between the particles and leads to a slightly higher collecting rate in this case (run E1).
}

\ownupdate{Using a speed of \SI{300}{rpm} (run E2), material can be collected.  However, material
that reaches the tray after \SI{0.25}{\s} can also fall back through the opening and the actual
collecting phase starts at \SI{0.5}{\s}, as soon as the number of particles in the tray rises
monotonically. At a speed of \SI{500}{rpm}, the material transport from the surface to the tray
becomes fast enough to prevent regolith in the tray from falling back down, and the number of
collected particles increases continuously.}

\ownupdate{In agreement with the experiments, we find that a rotation speed significantly lower than
\SI{500}{rpm} does not lead to mass transfer into the tray. \cite{astra2015} applied rotation rates
between \SIrange{500}{1000}{rpm} in their study in order to achieve sampling.}

\ownupdate{
\subsection{Sampling on Phobos}
}
We use the same
material properties and change the gravitational acceleration to a mean value on Phobos'
surface. The gravitational acceleration on its surface varies between
\refone{\SIrange{4e-3}{8e-3}{\metre \per \s^2}},
and we use a mean value of \SI{5.7e-3}{\metre \per \s^2}.

\begin{figure*}
    \begin{center}
        \includegraphics[width=0.8\textwidth]{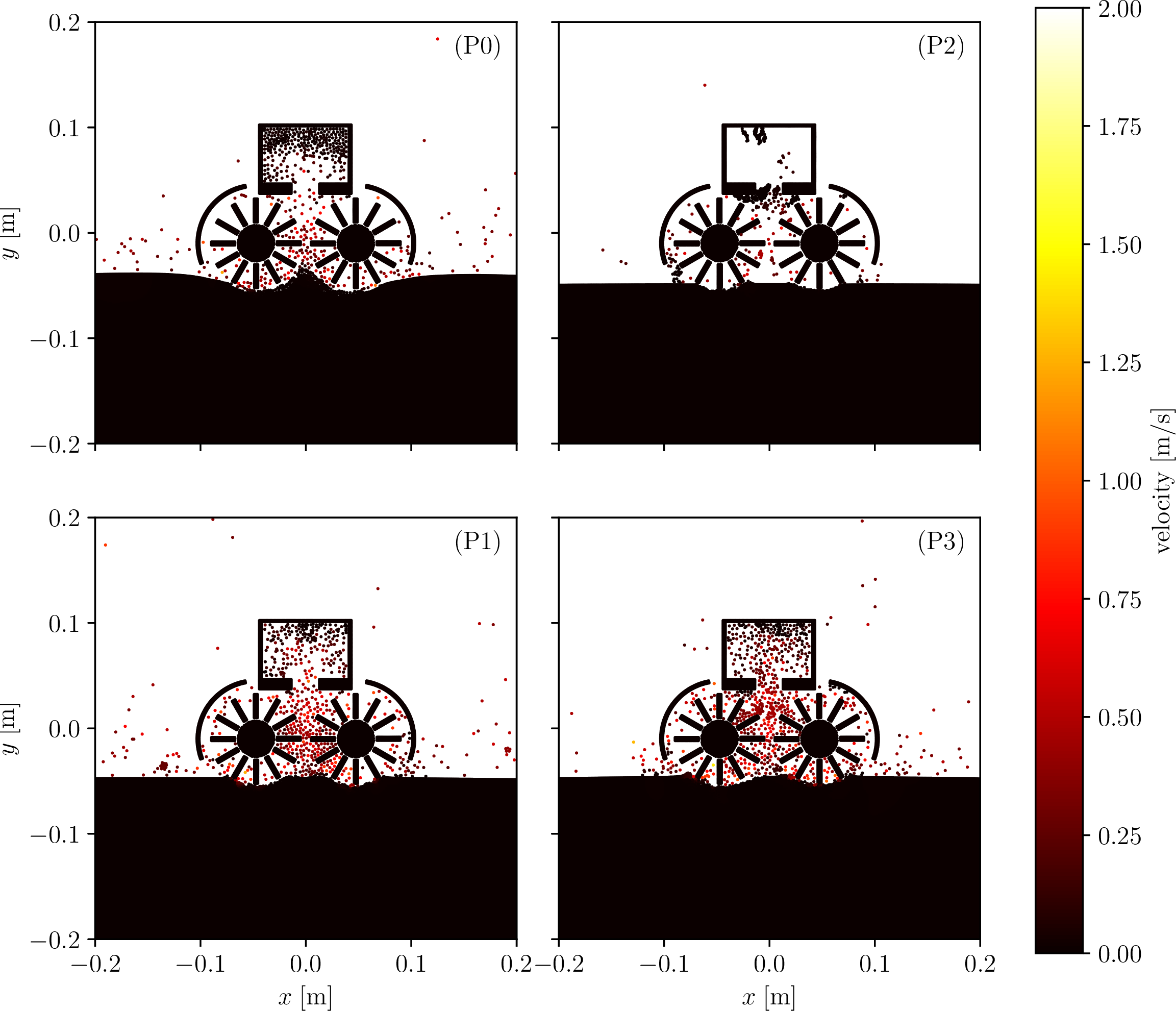}
    \end{center}
    \caption{Particle distribution after \SI{1}{\s} simulation time for four runs with Phobos'
    gravity. The velocity is color-coded. \reftwo{The rotation speed of the brushes is \SI{100}{rpm}
    for all four simulations. The angle of internal friction is \SI{42}{deg} for runs P0, P1 and P3
    and \SI{20}{deg} for P2. Run P0 is cohesionless, P1 with \SI{500}{\Pa}, and with \SI{1000}{\Pa}
    for runs P2 and P3.}
    \label{fig:phobos0}}
\end{figure*}

\begin{figure*}
    \begin{center}
        \includegraphics[width=0.8\textwidth]{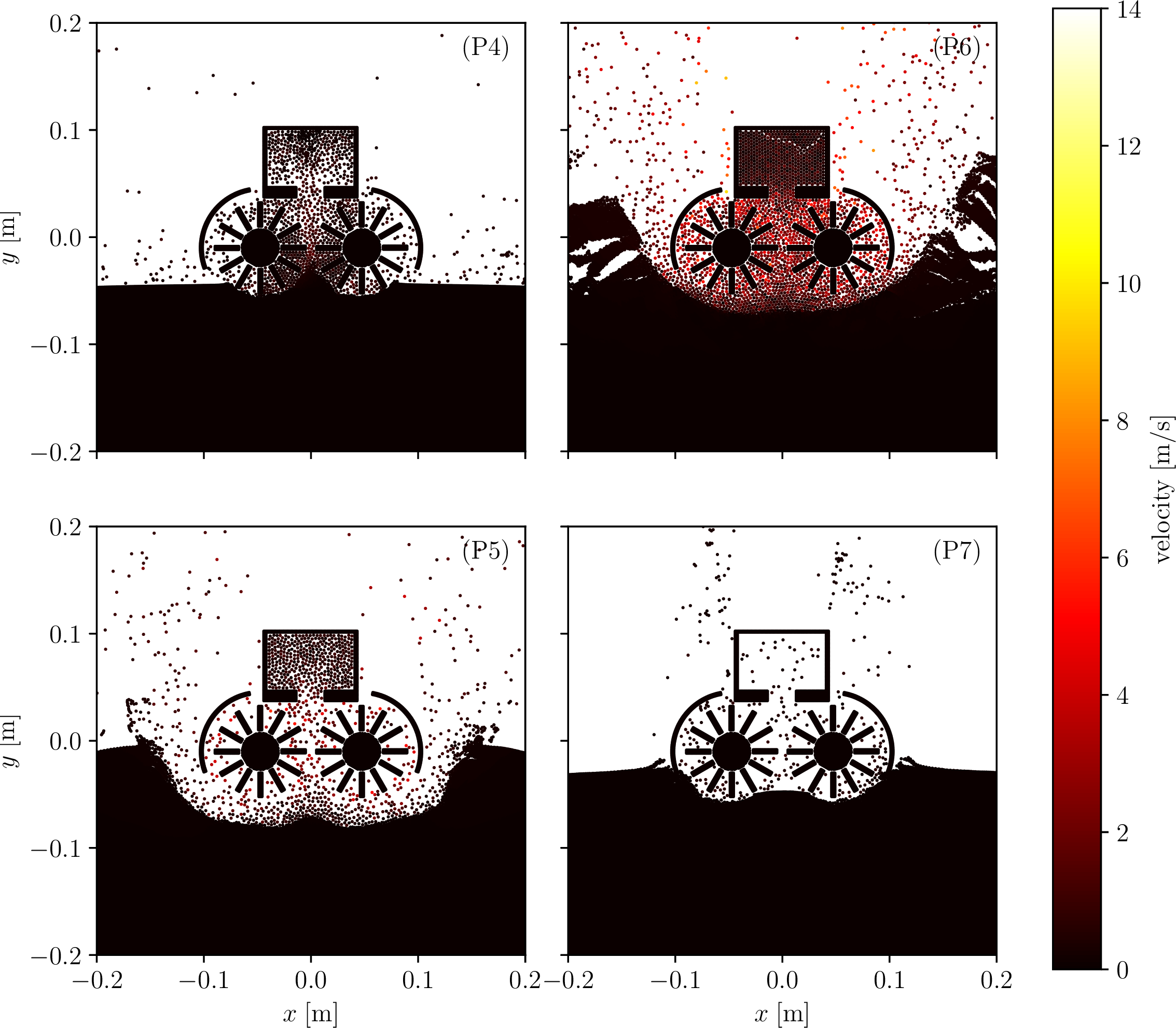}
    \end{center}
    \caption{Particle distribution after \SI{1}{\s} simulation time for four runs with Phobos'
        gravity. The velocity is color-coded. \reftwo{The runs have various different simulation
            parameters: Run P4 has an angle of friction of \SI{60}{deg} and runs P5, P6 and P7
            \SI{42}{deg}, runs P4 and P6 have a cohesion of \SI{1}{\kilo \Pa} and runs P5 and P7
            are cohesionless, the rotation speed of the brushes is
        \SIlist{100;500;500;-100}{rpm} for runs P4, P5, P6, and P7, respectively.}
    \label{fig:phobos1}}
\end{figure*}

\ownupdate{The positions of the particles after \SI{1}{\s} can be seen in the plots of
Figures~\ref{fig:phobos0} and \ref{fig:phobos1}. The colour code indicates the speed of the
particles. Runs P0-3 (Figure~\ref{fig:phobos0}) have the same brush rotation speed of
\SI{100}{rpm}, the material properties of the regolith were varied: The runs P0, P1 and P3 
include an angle of internal friction of \SI{42}{deg} and a cohesion of \SI{0}{\Pa}, \SI{500}{\Pa}
and \SI{1}{\kilo\Pa}, respectively, and P2 a lower angle of internal friction of \SI{20}{deg} and a
cohesion of \SI{1}{\kilo\Pa}. The runs with \SI{42}{deg} and cohesion show some small clumping of the
particles in the tray (P1 and P3). In the run with \SI{20}{deg} and \SI{1}{\kilo\Pa}, the opening of
the collecting tray gets blocked by a smaller pile, and material that entered the tray sticks
for the most part on the ceiling. 
}

\ownupdate{The rotation speed of the brushes was varied in runs P4-7 (Figure~\ref{fig:phobos1}): Runs
P5 and P6 have a rotation speed of \SI{500}{rpm}, P4 of \SI{100}{rpm} and P7 of \SI{100}{rpm} in the
opposite direction. Runs P5-7 include an angle of internal friction of \SI{42}{deg} and P4 of
\SI{60}{deg}. The runs P5 and P7 are without cohesion and runs P4 and P6 include a cohesion of
\SI{1}{\kilo\Pa}. Note the different velocity scaling compared to Figure~\ref{fig:phobos0}. Both runs
with the highest rotation rate (P5 and P6) show the largest impact on the surface. In the case with a
cohesion of \SI{1}{\kilo\Pa} (P6), the particles fill the space in the collecting tray and in between
the bristles of both brushes. On both sides, larger piles of regolith get pushed away. The same
simulation setup without any cohesion (P5) shows a less dramatic picture with slower velocities. The
cavity below the brushes is deeper than in the other runs and less particles stick in between the
bristles.}

\begin{figure}
    \begin{center}
        \includegraphics[width=0.5\textwidth]{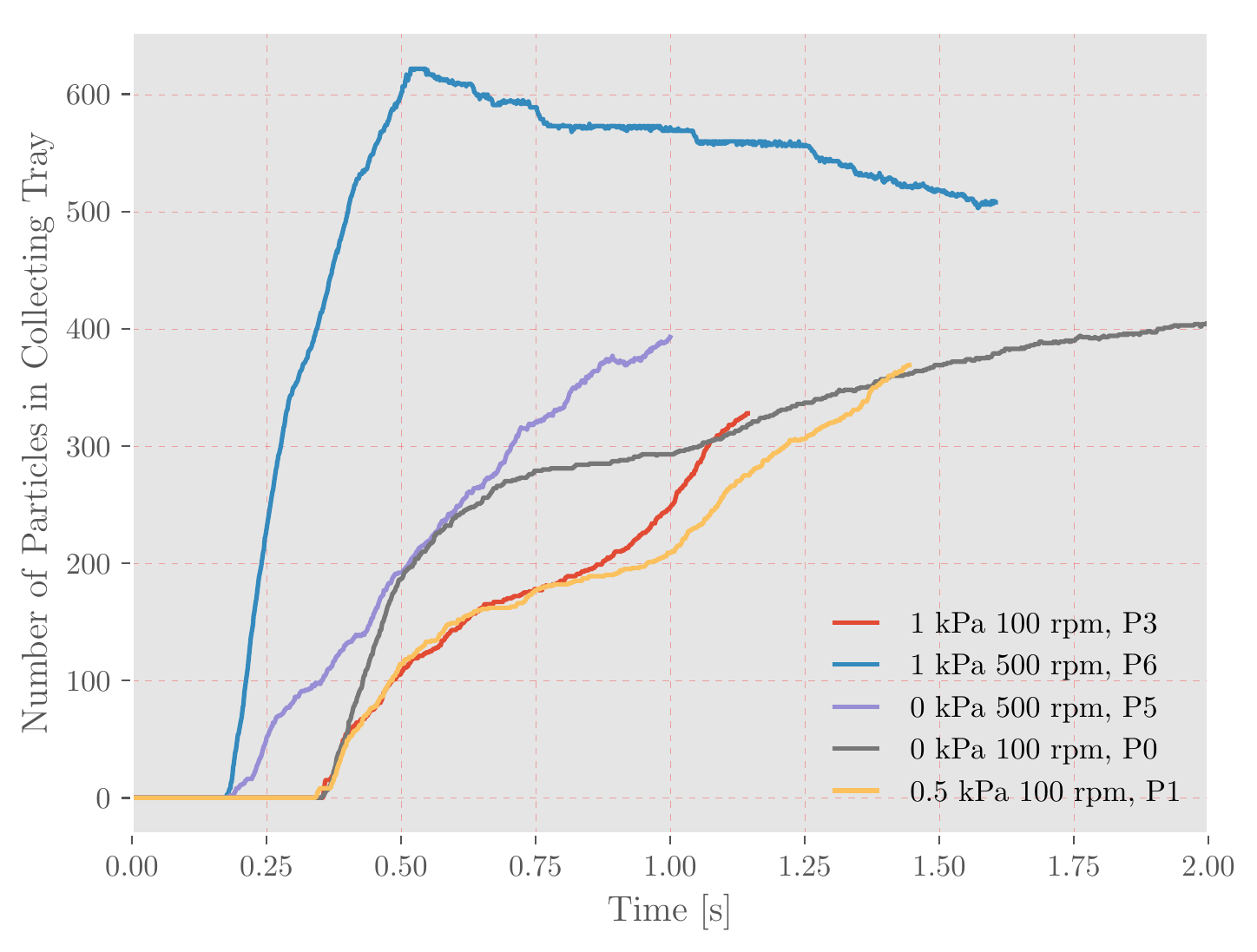}
    \end{center}
    \caption{Sampling process on Phobos' surface for two different rotation speeds of the brushes.
    The angle of internal friction of the regolith is \SI{42}{deg}. 
        \label{fig:dust_analysis_phobos42}}
\end{figure}

\ownupdate{To analyze the collecting process in more detail, the time evolution of the amount of
sampled material in the collecting tray on Phobos' surface is shown in
Figure~\ref{fig:dust_analysis_phobos42}. Higher rotation speeds yield higher sampling rates. In the
case of \SI{500}{rpm} and a cohesion of \SI{1}{kPa} (blue curve, run P6), the collecting tray is
filled after $\SI{0.5}{\s}$ after simulation start, upper right
panel), and subsequently the amount of regolith in the tray decreases.}

\subsubsection{Varying angle of internal friction and cohesion}
\begin{figure}
    \begin{center}
        \includegraphics[width=0.5\textwidth]{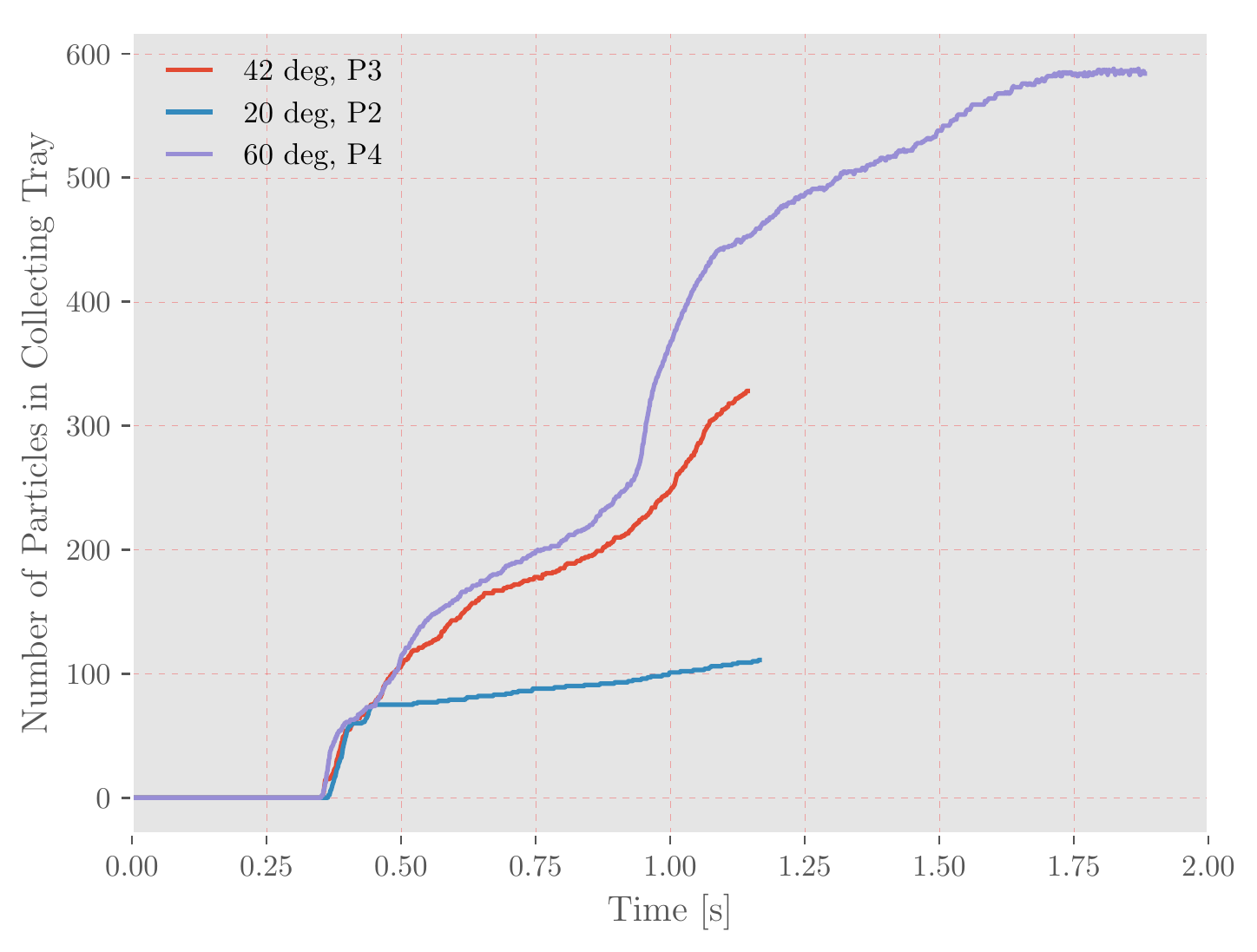}
    \end{center}
    \caption{Influence of the angle of internal friction. The collected number of particles for
    Phobos' gravity and cohesion of $\SI{1}{kPa}$ are shown for three different angles of friction. The
    rotation speed of the brushes is set to \SI{100}{rpm}. \label{fig:dust_analysis_phobos_angle}}
\end{figure}

The results for the three simulations with varying angle of internal friction on Phobos are shown in
Figure~\ref{fig:dust_analysis_phobos_angle}. 
\ownupdate{The internal friction leads to higher shear in
the regolith, the force between the particles raises, which eventually supports the sampling
process, since the brushes break off larger clumps at first contact with the surface.}
In the case of a small internal friction (P2), the
collecting tray is much less filled compared to the simulation with \SI{42}{deg} (P3). In this
particular simulation, we observe that one small clump of regolith sticks at the left end of the
opening and blocks it partially (see Figure~\ref{fig:phobos0}). 

A further increase of the internal friction leads to an even higher collection rate (P4). However,
the value \SI{60}{deg} is higher than the expected value for regolith on Phobos. The most probable
range of lunar soil friction angle is about \SIrange{30}{50}{deg} \citep{1972LPSC....3.3235M} and
Phobos' regolith is presumably in a range comparable to lunar and martian soil.

\ownupdate{The influence of the cohesion on the collecting rate is indistinct. On the one hand,
cohesion supports the effect of clumping and larger chunks of regolith are transported by the
brushes, while on the other hand cohesion also hinders the brushes to lift the chunks from the
surface. We see a lower collection rate in the Phobos runs with \SI{100}{rpm} rotation speed
for cohesion of \SI{500}{Pa} and \SI{1}{\kilo\Pa} compared to the cohesionless case for the first
second of the simulation, see curves P0, P1 and P3 in Figure~\ref{fig:dust_analysis_phobos42}. The
run with \SI{1}{\kilo\Pa} cohesion, however, yields a slightly larger collection rate than the run
with \SI{500}{\Pa}.} 

Using a five times higher rotation speed of \SI{500}{rpm} brings
particles to the tray already after \SI{0.19}{\s}. Interestingly, a higher cohesion in combination
with a higher rotation speed at \SI{500}{rpm} yields to a faster collection rate (run P6) compared to
no cohesion (run P5), whereas for a lower rotation speed of \SI{100}{rpm}, cohesion lowers the
collection rate (runs P0, P1, P3). This is caused by the fact that higher cohesion leads to larger
clumps of regolith which requires more energy to be lifted by the brushes as can be seen quantitatively
in Figure~\ref{fig:phobos0} (panels P0, P1, P3 with cohesion \SI{0}{\Pa}, \SI{500}{\Pa}, and
\SI{1}{\kilo\Pa}, respectively). 

In the case of the highest cohesion and highest rotation speed (run
P6), the tray reaches its maximum charging level already after \SI{0.51}{\s}. Subsequently, material
from the tray falls back through the opening until eventually the pressure in the tray and the
pressure in the space between the brushes balance. Since the material stream from the surface is not
unlimited as the brushes stop their vertical motion after \SI{0.2}{\s}, the number of particles in
the tray slowly decays until the end of the simulation.

\begin{figure}
    \begin{center}
        \includegraphics[width=0.5\textwidth]{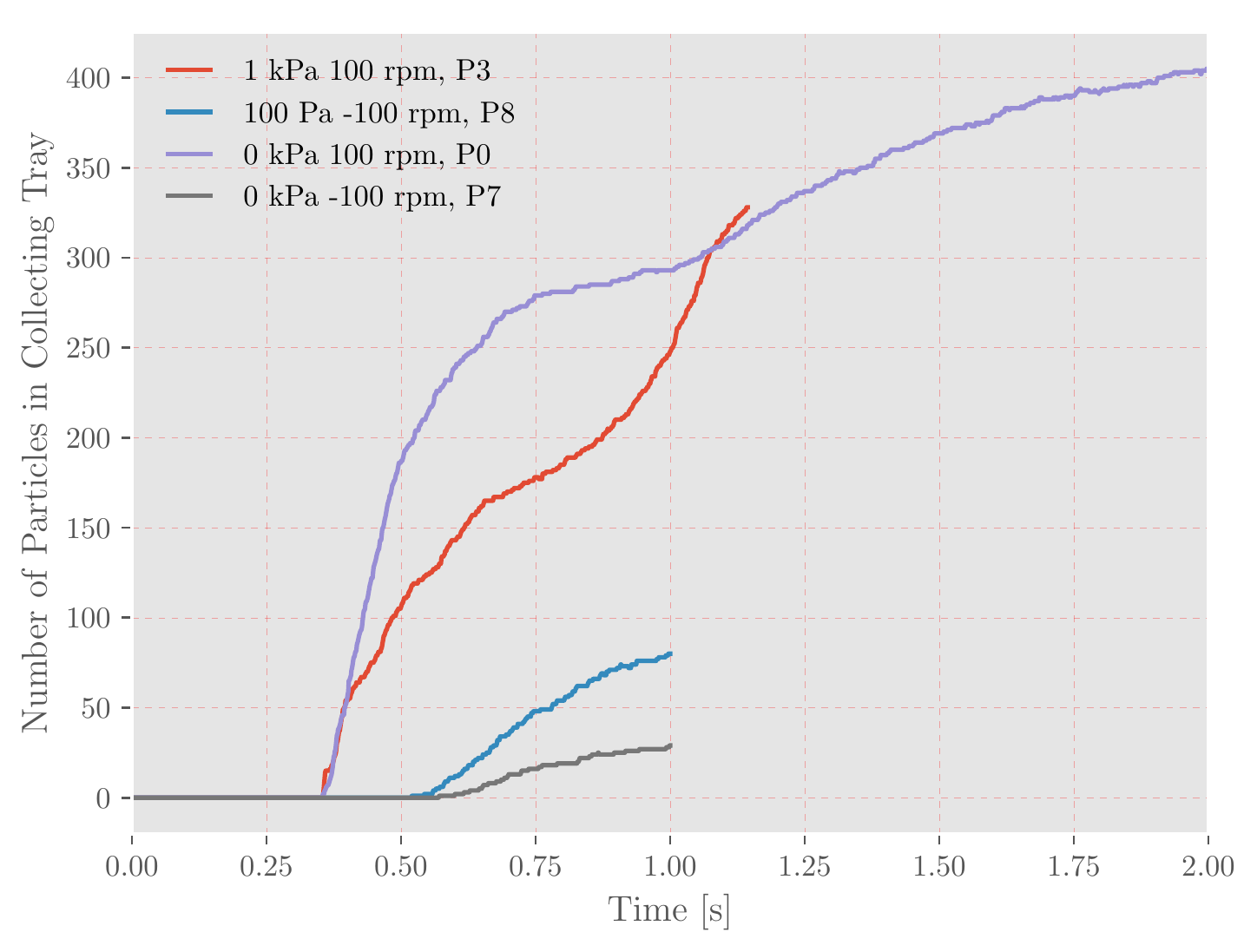}
    \end{center}
    \caption{Analysis of the clearing mechanism. 
    The sampling tool will additionally feature a clearing mechanism: The brushes rotate the
    other way than in the collecting process. In this manner, the surface below the brushes can be
    cleared of obstacles that might interfere during the collection phase. 
   %
    \label{fig:dust_analysis_counterclock}}
\end{figure}

\ownupdate{
\subsubsection{Clearing phase}
In the clearing phase, the brushes
rotate the other way.  In this manner, potential obstacles on the surface of Phobos may be cleared
and material from below the top layer can be collected in the subsequent collecting phase. We compare the collecting rate for a rotation
speed of \SI{100}{rpm} to the rotation speed of \SI{-100}{rpm} in
Figure~\ref{fig:dust_analysis_counterclock}, where 
the time evolution of the number of particles in the collecting tray for the clearing phase
simulation is shown.
In the first case, the first material
reaches the tray after about \SI{0.35}{\s}, independent from the cohesion. In the
latter case, in which no material at all should be collected, we still find small number of particles
that reach the tray.  These particles are transported by sticking between the bristles of each
brush. Their path to the collecting tray is longer than for the particles that are lifted between the
two brushes and they arrive later at the opening of the tray. The first particles enter the tray
after \SI{0.53}{\s}.} 

\begin{figure}
    \begin{center}
        \includegraphics[width=0.5\textwidth]{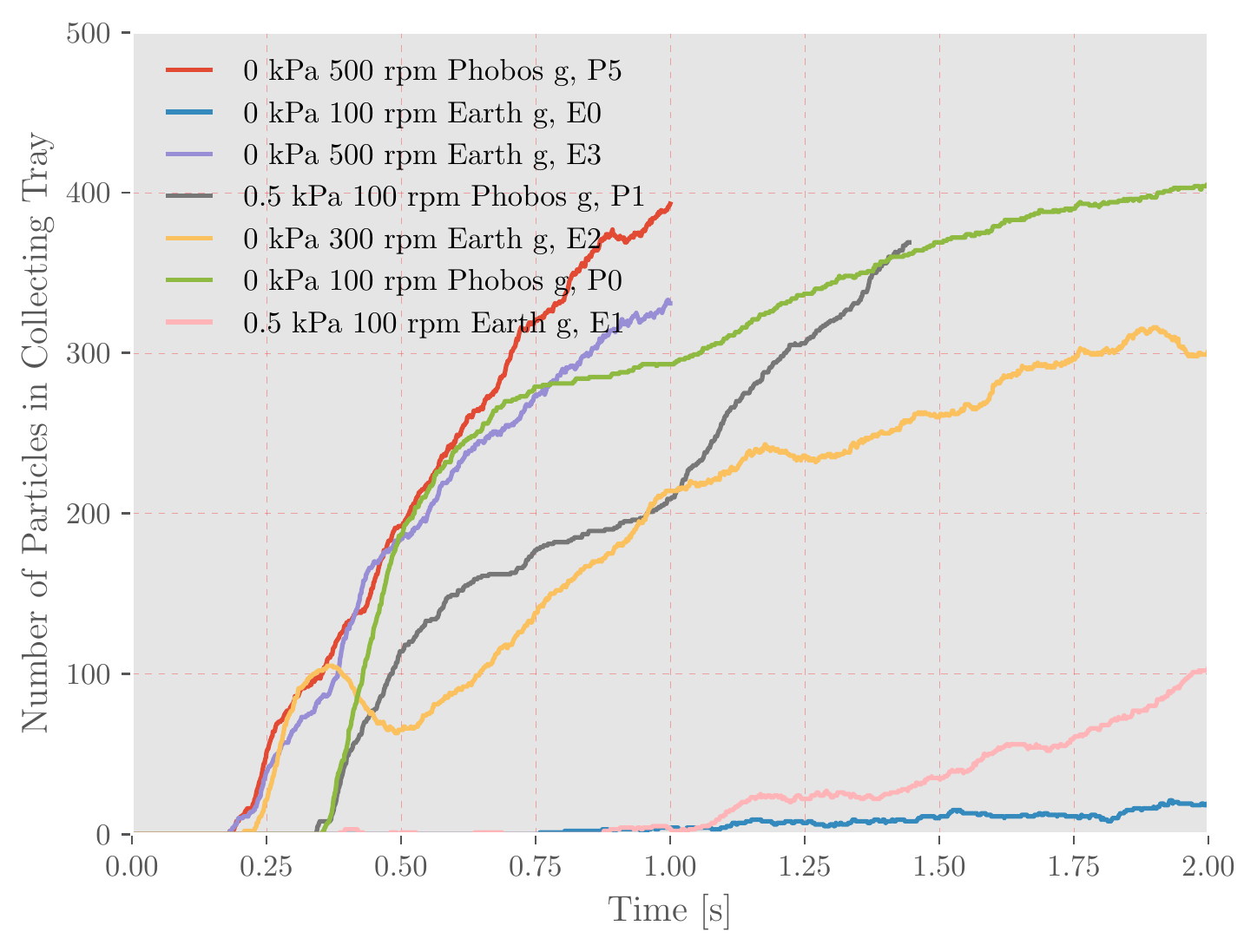}
    \end{center}
    \caption{Comparison between sampling on Earth and on Phobos and the influence of the surface gravity on
    the collecting process. The plot shows the collected number of particles for two different values
    of the cohesion, three different rotation speeds of the brushes and the surface gravity on Earth
    and Phobos. 
%
    \label{fig:dust_analysis_earth_phobos_comparison}}
\end{figure}

\ownupdate{
\subsection{Comparison Phobos vs.\ Earth}
}
\ownupdate{
The time evolution of the collected number of particles in direct comparison between Earth and Phobos
is shown in Figure~\ref{fig:dust_analysis_earth_phobos_comparison}.}

In contrast to the sampling on Earth, regolith can be transferred to the collecting tray on Phobos
even for the lowest investigated rotation speed of the brushes (\SI{100}{rpm}). At this speed, the first material
enters the tray after \SI{0.35}{\s} on Phobos (run P0), \ownupdate{compared to} \SI{0.75}{\s} on
Earth (run E0).  However, the sampling rate on Earth is \reftwo{significantly lower}, and after
\SI{1}{\s} there are less than 20 particles in the tray. \ownupdate{At this time, there are already
400 collected particles in run P0.}

%



%

%
%
%
%
\section{Conclusions\label{section:conclusions}} 
The simulations give some promising insight in\-to the pos\-sible application of SPH to regolith
modelling. The basic functionality of the sample process could be investigated for different material
properties of regolith. \refone{Since our model cannot provide absolute data for the collecting
phase, we compare the relative values between the individual simulations.}
\ownupdate{We have simulated the collection phase of the brush sampling tool for varying cohesion and
angles of internal friction on Earth and on Phobos.}

For non-cohesive soil,
Earth's gravity is high enough to prevent a high collecting rate of dust material at brushes speed
lower than \SIrange{100}{300}{rpm}. If the soil is slightly cohesive (about \SI{500}{Pa} cohesion),
the situation improves.  
For Phobos’ gravity and lunar soil properties (\SI{42}{deg}, \SI{500}{Pa}),
the collecting rate is more than ten times higher at \SI{100}{rpm} than for Earth's gravity. 

We found one case, in which the collecting tray hole was blocked by particles (\SI{20}{deg},
\SI{1}{kPa}, \SI{100}{rpm}, Phobos g, run P2) and the collecting rate was slowed down in the
following. This might come from the 2D geometrical setup and may not happen in 3D simulations.
However, to obtain a reasonable high spatial resolution, we chose a 2D geometrical setup.  The speed
of the dispersed dust particles is mainly determined by the rotation rate of the brushes and depends
only slightly on the material properties angle of internal friction and cohesion.  

The escape velocity of Phobos (\SI{11.39}{\metre \per \s}) was only reached in one simulation
(\SI{42}{deg}, \SI{1}{kPa}, \SI{500}{rpm}, run P6) after \SI{0.6}{\s}. However, this resulted from
some escaping particles that were pressed through the slit between the collecting tray and the shield
of the brushes.  This slit is not in the CAD data and only in the SPH setup. 

The shielding around the brushes is very effective: We do not see any material flying directly
upwards to the spacecraft.  Qualitatively, the simulations do not show high dust contamination of the
\reftwo{spacecraft} from the sampling process.  The maximum settling time of the dispersed dust in
our simulations can be estimated to be between \SIrange{40}{80}{minutes} (for Phobos g varying from
\SIrange{4e-3}{8e-3}{\metre \per \s^2}).

During the clearing (brushes turning the other way round), we find some particles that get eventually
in the collecting tray. These collected particles are transported within one single brush only
\reftwo{and enter the tray after \SI{0.53}{\s}. Cohesion enhances this effect since it supports the
sticking of material within the bristles of a single brush.  Therefore, it is recommended to keep the
tray closed during the clearing phase, and additionally a process to clean out the material in the
bristles of the brushes after the cleaning phase may be required to avoid material from the clearing
phase to be sampled.}

Our analysis is based on a series of simulations that do not cover the entire parameter space.
\ownupdate{It is desirable to broaden the parameters even more and additionally vary the density of the regolith
or add a layered structure to the surface model.}
Based on our experience with this study, we suggest following further steps to improve the sampling
model simulations
\begin{itemize}
\item Subsequent simulations of the collecting process need higher spatial resolution and should be
    three-dimensional.  Especially, the latter requirement is obvious if one wants to calculate
        absolute values for the collection rate, since the opening at the bottom of the collecting tray
        cannot be modelled correctly in two dimensions.
\reftwo{Recently, a study about the Mars Exploration Rover wheel performance was published by
    \cite{Johnson201531}. They successfully applied the three-dimensional discrete element method
(DEM) to investigate wheel slip conditions for various contact friction coefficients. They underline
the need for three-dimensional simulations in their study to accomplish good agreement with
experimental data.}

\item Simulations with different angles between sampling brushes and surface: The simulations for the
    sampling process presented in this work feature brushes, \reftwo{where the axes of the brushes} are oriented parallel to Phobos’
        surface and \reftwo{move only up and down}. Subsequent simulations may be more
        realistic if the sampling tool moves with more degrees of freedom. 
        Furthermore, we have
        simulated only one path of \reftwo{the brush axes}: from the starting point going with constant speed in vertical
        direction for \SI{0.2}{\s} and halting. A proper model of the brushes may include a feedback
        reaction of the brushes depending on the sampling rate.
        \ownupdate{The prototype by \cite{astra2015} had the additional possibility to manually
        adjust the angle between the surface of the regolith and the brushes with three different
        values. This could be added quite simply to our initial setup for a future study.}
\item  In this study, we have applied the soil model presented by \cite{NAG:NAG688}.  In a
        further study, we will compare this model to another model for granular media using
        continuous mechanics, where the soil can be treated as a viscous material \citep{Ulrich2013}.
\end{itemize}

In summary, in this study we have successfully applied the soil model by \cite{NAG:NAG688} for
regolith and built a simulation testbed for modeling the sampling process considered for PhSR. 
Although absolute values for the sampling rate cannot be provided, the dependencies of the
collecting process on various material properties such as the cohesion and angle of internal friction
can be studied in detail. 

Moreover, \reftwo{in agreement with the laboratory experiments with the sampling prototype by
\cite{astra2015}, we find that a rotation speed of \SI{100}{rpm} for the brushes is not sufficient
to transfer regolith from Earth's surface into the tray.} 

\ownupdate{In this study} we cannot provide absolute numbers for
sampling rates, and give estimations about the influence of the mechanical properties of the
regolith on the sampling rate. Using an improved three dimensional model and higher spatial
resolution, \ownupdate{future} numerical models will allow for more precise conclusions.

\section*{Acknowledgements}
This material is based on the work performed on behalf of the industrial Phobos Sample Return Phase A
study team of Airbus Defense and Space under ESA contract 4000114302/15/NL/PA (technical officer:
Thomas Voirin).
\ownupdate{We thank Gerhard Wurm and an anonymous reviewer whose comments and suggestions helped
to improve and clarify this manuscript.}\newline
All plots in this publication have been made by the use of the matplotlib package by
\cite{Hunter:2007}. \newline
TM appreciates support by the FWF Austrian Science Fund project S~11603-N16.
CS wants to thank Daniel Thun for helpful discussions during the course of this project.
\bibliographystyle{elsarticle-harv} 
\bibliography{phobos_paper}

\end{document}